\documentclass[12pt,preprint]{aastex}
\usepackage{psfig,natbib}

\usepackage{palatcm}

\def\la{\mathrel{\hbox{\rlap{\hbox{\lower4pt\hbox{$\sim$}}}\hbox{$<$}}}}
\def\ga{\mathrel{\hbox{\rlap{\hbox{\lower4pt\hbox{$\sim$}}}\hbox{$>$}}}}

\newcommand{\be}{\begin{eqnarray}}
\newcommand{\ee}{\end{eqnarray}}

\newcommand{\msol}{\ifmmode{{\rm M}_\odot}\else{M$_\odot$}\fi}
\newcommand{\foe}{\ifmmode{10^{51}}\else{$10^{51}$}\fi}

\newcommand{\xni}{\ifmmode{{\rm X}_{\rm Ni}}\else{X$_{\rm Ni}$}\fi}

\def\Teff{\ifmmode{T_{\rm eff}}\else{\hbox{$T_{\rm eff}$} }\fi}
\def\Rzero{\ifmmode{R_0}\else{\hbox{$R_0$} }\fi}

\def\SP2{{\tt IBM SP2}}

\def\PC2{{\tt PC$^2$}}

\def\inu{\ifmmode{I_{\nu}}\else{\hbox{$I_{\nu}$} }\fi}
\def\snu{\ifmmode{S_{\nu}}\else{\hbox{$S_{\nu}$} }\fi}
\def\jnu{\ifmmode{J_{\nu}}\else{\hbox{$J_{\nu}$} }\fi}

\def\fep{\ifmmode{{\rm Fe II}}\else\hbox{Fe~II }\fi}

  at 12.0 true pt 

\def\phoenix{{\tt PHOENIX}}

\def\water{{H$_2$O}}

\def\phoenix{{\tt PHOENIX}}

\def\water{{H$_2$O}}

\def\b{\beta}

\def\rout{\ifmmode{r_{\rm out}}\else\hbox{$r_{\rm out}$}\fi}
\def\tmax{\ifmmode{\tau_{\rm max}}\else\hbox{$\tau_{\rm max}$}\fi}
\def\tstd{\ifmmode{\tau_{\rm std}}\else\hbox{$\tau_{\rm std}$}\fi}
\def\vmax{\ifmmode{v_{\rm max}}\else\hbox{$v_{\rm max}$}\fi}
\def\muE{\ifmmode{\mu_{\rm E}}\else\hbox{$\mu_{\rm E}$}\fi} 
\def\pE{\ifmmode{p_{\rm E}}\else\hbox{$p_{\rm E}$}\fi} 
\def\bmax{\ifmmode{\b_{\rm max}}\else\hbox{$\b_{\rm max}$}\fi}

\def\Teff{\hbox{$\,T_{\rm eff}$} }

\def\rout{\hbox{$r_{\rm out}$} }

\def\chistd{\ifmmode{\chi_{\rm std}}\else\hbox{$\chi_{\rm std}$}\fi}

\def\msol{$M_\odot$}
\def\foe{10^{51}}

\def\xni{{\rm X}_{\rm Ni}}

\def\lstar{\ifmmode{\Lambda^*}\else\hbox{$\Lambda^*$}\fi} 
\def\Rop{\ifmmode{[R_{ij}]}\else\hbox{$[R_{ij}]$}\fi}

\def\Rji{\ifmmode{[R_{ji}]}\else\hbox{$[R_{ji}]$}\fi}
\def\Rstar{\ifmmode{[R_{ij}^*]}\else\hbox{$[R_{ij}^*]$}\fi}

\def\Rjistar{\ifmmode{[R_{ji}^*]}\else\hbox{$[R_{ji}^*]$}\fi}
\def\DRji{\ifmmode{[\Delta R_{ji}]}\else\hbox{$[\Delta R_{ji}]$}\fi}
\def\DRij{\ifmmode{[\Delta R_{ij}]}\else\hbox{$[\Delta R_{ij}]$}\fi}

\def\ns{\ifmmode{N_{\rm s}}          
        \else\hbox{$N_{\rm s}$}\fi}

\def\mat#1{{\bf #1}}     
\def\vek#1{{#1}}         

\newcount\eqcount
\def
  \nummer{
    \global\advance\eqcount by 1
    (\the\eqcount)
  }

\def
  \numadv{
    \global\advance\eqcount by 1
  }

\def
   \numout#1{
     (\the\eqcount #1)
  }

\def\ivek#1#2{\ifmmode{\vek{I}^{#1}_{#2}}
        \else\hbox{$\vek{I}^{#1}_{#2}$}\fi}

\def\tmat#1#2{\ifmmode{\mat{t}^{#1}_{#2}}
        \else\hbox{$\mat{t}^{#1}_{#2}$}\fi}
\def\rmat#1#2{\ifmmode{\mat{r}^{#1}_{#2}}
        \else\hbox{$\mat{r}^{#1}_{#2}$}\fi}
\def\bvek#1#2{\ifmmode{\beta^{#1}_{#2}}
        \else\hbox{$\beta^{#1}_{#2}$}\fi}

\def\lp{\ifmmode{\lambda^+_\tau}           
        \else\hbox{$\lambda^+_\tau$}\fi}
\def\lm{\ifmmode\lambda^-_\tau             
        \else\hbox{$\lambda^-_\tau$}\fi}

\chardef\tilt=126
\begin{document}

\bibliographystyle{apj}

\title{Parallel Implementation of the {\tt PHOENIX} Generalized Stellar 
Atmosphere Program. III: A parallel algorithm for direct opacity sampling}

\author{Peter H. Hauschildt}
\affil{Dept.\ of Physics and Astronomy \& Center for Simulational Physics, 
University of Georgia, Athens, GA 30602-2451\\
Email: {\tt yeti@hal.physast.uga.edu}}
\author{David K. Lowenthal}
\affil{Dept.\ of Computer Science,
University of Georgia, Athens, GA 30602\\
Email: {\tt dkl@cs.uga.edu}}
\author{E.~Baron}
\affil{Dept. of Physics and Astronomy, University of
Oklahoma, 440 W.  Brooks, Rm 131, Norman, OK 73019-0225;\\
Email: {\tt baron@phyast.nhn.ou.edu}}

\begin{abstract}

We describe two parallel algorithms for line opacity calculations based on a
local file and on a global file approach. The performance and scalability of
both approaches is discussed for different test cases and very different
parallel computing systems. The results show that a global file approach is
more efficient on high-performance parallel supercomputers with dedicated
parallel I/O subsystem whereas the local file approach is very useful on farms
of workstations, e.g., cheap PC clusters.

\end{abstract}

\section{Introduction}


In the first 2 papers of this series \cite[][hereafter paper I and II,
respectively]{parapap,parapap2} we have described general parallel algorithms
that we have implemented in our multi-purpose stellar atmosphere code \phoenix.
These papers focused mainly on radiative transfer and NLTE problems and general
parallelization issues. In this paper we discuss in greater depth the problem
of line opacity calculations. This is in particular important if extensive line
lists are used, i.e., molecular line databases. These databases have increased
dramatically over the last decade, mostly due to the work of
\cite{UCL,schryb94,ames-water-new} on water vapor lines and
\cite{TiOJorg,ames-tio} on TiO lines. Currently our molecular line database
contains about 550 million lines, most of which are TiO and \water\ lines. 
Using these databases for opacity calculations poses a significant challenge
both for the construction of opacity tables and for the construction of detailed
model atmospheres.

In our model atmosphere code we have implemented and used direct opacity
sampling (dOS) for more than a decade with very good results. During that
time, the size of the combined atomic and molecular line databases that we used
has increased from a few MB to $>8\,$GB. Whereas the floating point and memory
performance of computers has increased dramatically in this time, I/O
performance has not kept up with this speed increase.
Presently, the wallclock times used by the  line selection and opacity
modules are dominated by I/O time, not by floating point or overall 
CPU performance.
Therefore, I/O
performance is today more important that it was  10 years ago and has to be
considered a major issue. The availability of large scale parallel
supercomputers that have effectively replaced vector machines in the last 5
years, has opened up a number of opportunities for improvements of dOS
algorithms. Parallel dOS algorithms with an emphasis on the handling of large
molecular line databases are thus an important problem in computational stellar
atmospheres. These algorithms have to be portable and should perform well for
different types of parallel machines, from cheap PC clusters using Ethernet
links to high performance parallel supercomputers. This goal is extremely hard
to attain on all these different systems, and we, therefore, consider two
different parallel dOS algorithms in this paper and compare their performance
on two very different parallel machines. In the next sections we will discuss
the direct opacity sampling method, describe the parallel algorithms in detail
and then discuss the results of test calculations. We close with a summary
and conclusions.

\section{Direct Opacity Sampling}

There are a number of methods in use to calculate line opacities. The classical
methods are statistical and construct tables that are subsequently used in the
calculation. The Opacity Distribution Function (ODF) and its derivative the
k-coefficient method have been used successfully in a number of atmosphere and
opacity table codes \cite[e.g.,][]{kurucz92}. This method works well for
opacity table  and  model construction but cannot be used to calculate detailed
synthetic spectra. A second approach is the opacity sampling (OS) method
\cite[e.g.,][]{peytremann74}. This is a statistical approach in which the line opacity
is sampled on a fine grid of wavelength points using detailed line profiles for
each individual spectral line. In classical OS implementation, tables of
sampling opacities are constructed for given wavelengths grids and for
different elements.  These OS tables are then used to calculate model
atmospheres and, e.g., Rosseland mean opacities. The OS method has the
advantage that is more flexible than the ODF approach and it also allows the
construction of (typically) low resolution synthetic spectra. The drawback of
tables in general is, however, an inherent inflexibility in terms of, e.g., the
wavelength grid or the tables's resolution. For example, to properly account
for the pressure broadening of lines an opacity sampling table would have to be
a function of temperature and gas pressure, which leads to very large tables if
many wavelength points are tabulated (this is not a problem for ODF tables as
the number of wavelength bins in such models is typically very small). In
addition, a different code is typically required to calculate high-resolution
spectra from the model atmosphere constructed with the ODF or OS tables, which
has the potential of introducing systematic errors (e.g., if the
atmosphere/table and the synthetic spectrum codes are not synchronized).

In direct opacity sampling (dOS) these problems are avoided by replacing the
tables with a direct calculation of the total line opacity at each wavelength
point for all layers in a model atmosphere \cite[e.g.][]{jcam}. In the dOS
method the relevant lines (defined by a suitable criterion) are first selected
from master spectral line databases which include all available lines. The line
selection procedure will typically select more lines than can be stored in
memory and thus temporary line database files are created during the line
selection phase. The file size of the temporary database can vary, in theory,
from zero to the size of the original database or larger, depending on the
amount of data stored for the selected lines and their number. For large
molecular line databases this can easily lead to temporary databases of 
several GB in size. This is in part due the storage for the temporary line
database: its data are stored for quick retrieval rather than in the compressed
space saving format of the master line databases. The number and identity of lines that 
are selected from the master databases depends on the physical conditions
for which the line opacities are required (temperatures, pressures, abundances
for a model atmosphere) and thus the line selection has to be repeated if 
the physical conditions change significantly. As an optimization, it is 
easily possible to include only lines in the temporary database that can 
be ``seen'' by the wavelength grid that will be used in the calculation
of the line opacities later on. This is important if, for example, only 
a narrow range in wavelength is considered at high resolution in order to
generate a synthetic spectrum.

The temporary line databases are used in the next phase to calculate the actual
line opacity for each wavelength point in a prescribed (arbitrary) wavelength
grid. This makes it possible to utilize detailed line profiles for each
considered spectral line on arbitrary wavelength grids. For each wavelength
grid point, all (selected) lines within prescribed search windows (large enough
to include all possibly important lines but small enough to avoid unnecessary
calculations) are included in the line opacity calculations for this wavelength
point. This procedure is thus very flexible, it can be used to calculate line
opacities for both model atmosphere construction (with relatively few wavelength
points) and for the generation of high-resolution synthetic spectra. Its main
drawback is that the line selection and (in particular) line profile
calculations are more costly than table interpolations. 

\section{Parallel Algorithms}


There are currently a large number of significantly different types of parallel
machines in use, ranging from clusters of workstations or PCs to teraflop
parallel supercomputers. These systems have very different performance
characteristics that need to be considered in parallel algorithm design.  For
the following discussion we assume this abstract picture of a general parallel
machine: The parallel system consists of a number of processing elements (PEs),
each of which is capable of executing the required parallel codes or sections
of parallel code. Each PE has access to (local) memory and is
able to communicate data with other PEs through a communication
device. 
The PEs have access to both local and global filesystems for data
storage.
The local filesystem is 
private to each PE (and inaccessible to other PEs), and the global
filesystem can be accessed by any PE.
A PE can be a completely independent computer such as a PC or
workstation 
(with single CPU, memory, and disk),
or it can be a part of a shared memory multi-processor system.
For the purposes of this paper, we assume that the parallel machine has both
global and local logical filesystem storage available (possibly on the same
physical device). The communication device could be realized, for example, by
standard Ethernet, shared memory, or a special-purpose high speed
communication network.

In the following description of the 2 algorithms that we consider here
we will make use of the following features of the line databases:
\begin{itemize}
\item master line databases:
\begin{enumerate}
\item are globally accessible to all PEs 
\item are are sorted in wavelength
\item can be accessed randomly in blocks of prescribed size (number of lines)
\end{enumerate}
\item selected line (temporary) databases:
\begin{enumerate}
\item the wavelength grid is known during line selection (not absolutely required but helpful)
\item have to be sorted in wavelength
\item can be accessed randomly in blocks of prescribed fixed size (number of lines)
\item are stored either globally (one database for all PEs) or locally (one for each PE)
\item are larger than the physical memory of the PEs
\end{enumerate}
\end{itemize}

\subsection{Global Temporary Files (GTF)}


The first algorithm we describe relies on global temporary databases for the
selected lines. This is the algorithm that was implemented in the versions of
\phoenix\ discussed in papers I and II. In the general case of $N$ available
PE's, the parallel line selection algorithm uses one PE dedicated to I/O and $(N-1)$ line
selection PEs. The I/O PE receives data for the selected lines from the line
selection PEs, assembles them into properly sorted blocks of selected
lines,
and
writes them into the temporary database for later retrieval. The $(N-1)$ line
selection PEs each read one block from a set of $(N-1)$ adjacent blocks of line
data from the master database, select the relevant lines, and send the necessary
data to the I/O PE. Each line selection PE will select a different number of
lines, so the I/O PE has to perform administrative work to construct sorted
blocks of selected lines that it then writes into the temporary database. The
block sizes for the line selection PEs and for the temporary database created
by the I/O PE do not have to be equal, but can be chosen for convenience.  The
blocks of the master line database are distributed to the $(N-1)$ line
selection PEs in a round robin fashion. Statistically this results in a
balanced load between the line selection PEs due to the physical
properties of the line data.

After the line selection phase is completed, the temporary global line database
is used in the line opacity calculations. If each on the $N$ PEs is
calculating line opacities (potentially for different sets of wavelengths
points or for different sets of physical conditions), they all access the
temporary database simultaneously, reading blocks of line data as required. In
most cases of practical interest, the same block of line data will be accessed
by several (all) PEs at roughly the same time. This can be advantageous or
problematic, depending on the structure of the file system on which the
database resides. The PEs also cache files locally (both through the operating
system and in the code itself through internal buffers) to reduce
explicit disk I/O. Note that during the line selection phase the temporary
database is a write-only file, whereas during the opacity
calculations the temporary database is strictly read-only.

The performance of the GTF algorithm depends strongly on
the performance of the global file system used to store the temporary 
databases and on the characteristics of the individual PEs. This issue is discussed
in more detail below.

\subsection{Local Temporary Files (LTF)}



The second algorithm we consider in this paper uses file systems that are local
to each PE; such local file systems exist on many parallel machines, including
most clusters of workstations.  This algorithm is a new development discussed
here for the first time that tries to take advantage of fast communication
channels available on parallel computers and utilized local disk space space
for temporary line list files.  This local disk space is frequently large
enough for the temporary line database and may have high local I/O performance.
In addition, I/O on the local disks of a PE does not require any inter-PE
communication, whereas globally accessible filesystems often use the same
communication channel that explicit inter-PE communication uses. The latter can
lead to network congestion if messages are exchanged simultaneously with global
I/O operations.


For the line selection, we could use the algorithm described above with the
difference that the I/O PE would create a global (or local) database of
selected lines.  After the line selection is finished, the temporary database
could then simply be distributed to all PEs, and stored on their local disk for
subsequent use.  However, this is likely to be slower in all cases than the GTF
algorithm.

Instead, we use a ``ring'' algorithm that creates the local databases directly.
In particular, each of the $N$ PEs selects lines for one block from the master
database (distributed in round robin fashion between the PEs).  After the
selection for this one block is complete, each PE sends the necessary data to
it next neighbor; PE $i$ sends its results to PE $i+1 {\rm mod\;} N$ 
sends to PE 0) and, simultaneously, receives data from the previous PE in the
ring.  This can be easily realized using the MPI \cite[]{mpistd} {\tt
mpi\_sendrecv} call, which allows the simultaneous sending and receiving of data
for each member of the ring.  Each PE stores the data it receives into a buffer
and the process is repeated until the all selected lines from the $N$ blocks
are buffered in all $N$ PEs.  The PEs then transfer the buffered line data into
their local temporary databases. This cycle is repeated until the line
selection phase is complete. The line opacity calculations will then proceed in
the same way as outlined above, however, the temporary line databases are now
local for each PE. 

This approach has the advantage that accessing the temporary databases 
does not incur any (indirect) Network File System (NFS) communication
between the PEs as each of 
them has its own copy of the database. However, during the line selection
phase a much larger amount of data has to be communicated over the
network between the 
PEs because now each of them has to ``know'' all selected lines, not only
the I/O PE used in the first algorithm.

The key insight here is that low-cost parallel computers constructed
out of commodity workstations typically have a very fast communication
network (100 Mbs to 1 Gbs) but relatively slow NFS performance.
This means that trading off the extra communication for fewer NFS
disk accesses in the LTF algorithm is likely to
give better performance.

\section{Results}

The performance of the GTF and LTF algorithms will depend strongly on the 
type of parallel machine used. 
A machine using NFS with fast local disks
and communication 
is likely to perform better with the LTF algorithm. However,
a system with fast (parallel) filesystem and fast communication can actually
perform better with the GTF algorithm. 
In the following we will consider
test cases run on very different machines:
\begin{enumerate}
\item A cluster of Pentium Pro 200MHz PCs with 64MB RAM, SCSI disks, 100Mbs full-duplex 
Ethernet communication network, running Solaris 2.5.1.
\item An IBM SP system with 200MHz Power3 CPUs, 512MB RAM per CPU, 133 MB/s switched
communication network, 16 node IBM General Purpose File System (GPFS) parallel 
filesystem, running AIX 4.3.
\end{enumerate}


We have run 2 test cases to analyze the behavior of the 
different algorithms on different machines. The small test
case was designed to execute on the PPro system. It uses
a small line database (about 550MB) with about 35 million lines
of which about 7.5 million lines are selected. The second
test uses an about 16 times larger line database (about 9GB)
and also selects 16 times more lines. This large test
could not be run on the PPro system due to file size 
limitations and limited available disk space. The line
opacity calculations were performed for about 21,000 
wavelength points that are representative for typical 
calculations. The tests were designed for maximum I/O 
usage and are thus extreme cases. In practical
applications the observed scaling is comparable to or
better than which found for these tests and appears to 
follow the results shown here rather well.
In the following we will discuss
the results for the tests on the different computing systems.

\section{PPro/Solaris system}

\begin{figure}[b]
\psfig{file=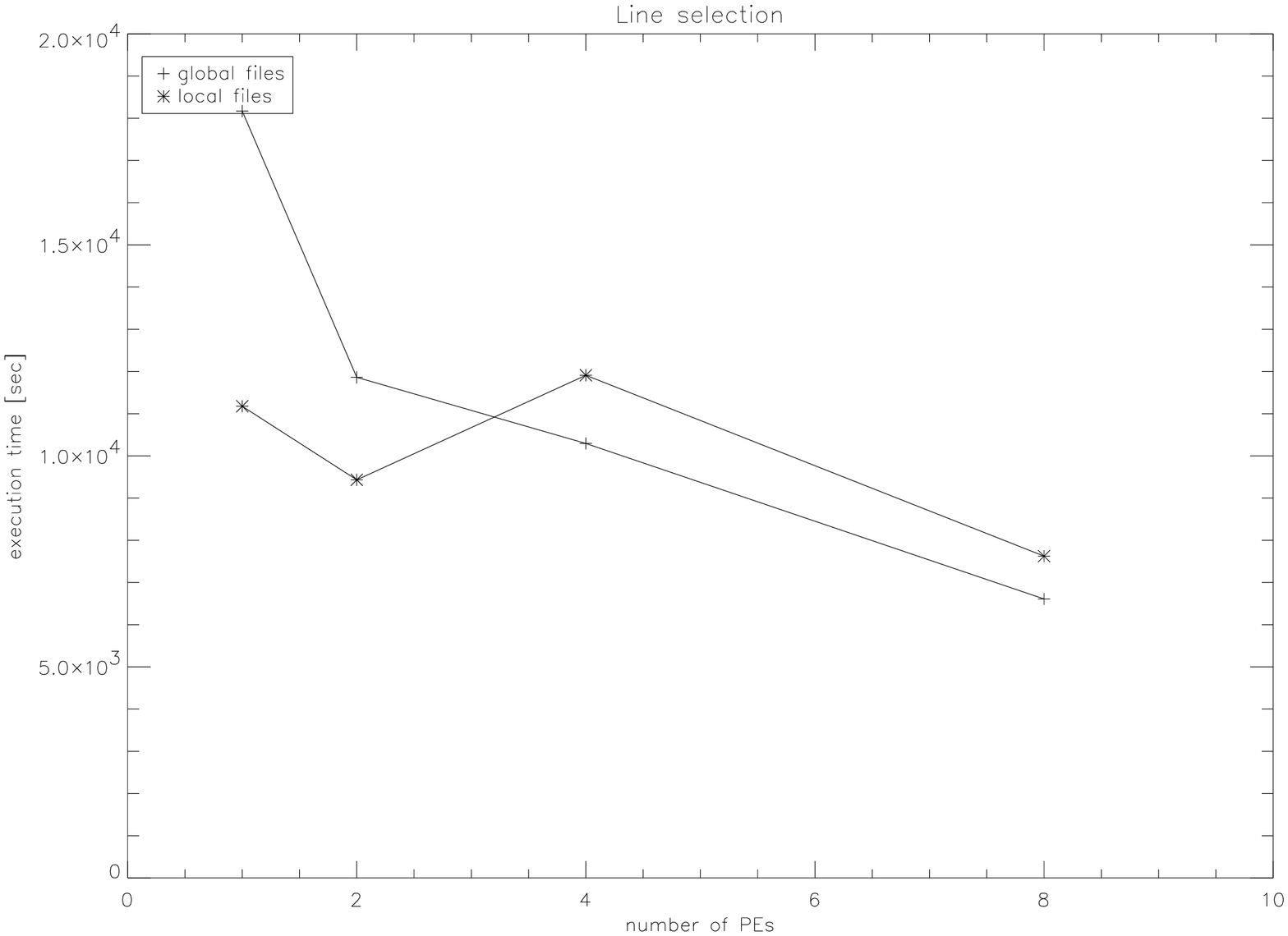,width=0.75\hsize,angle=90}
\psfig{file=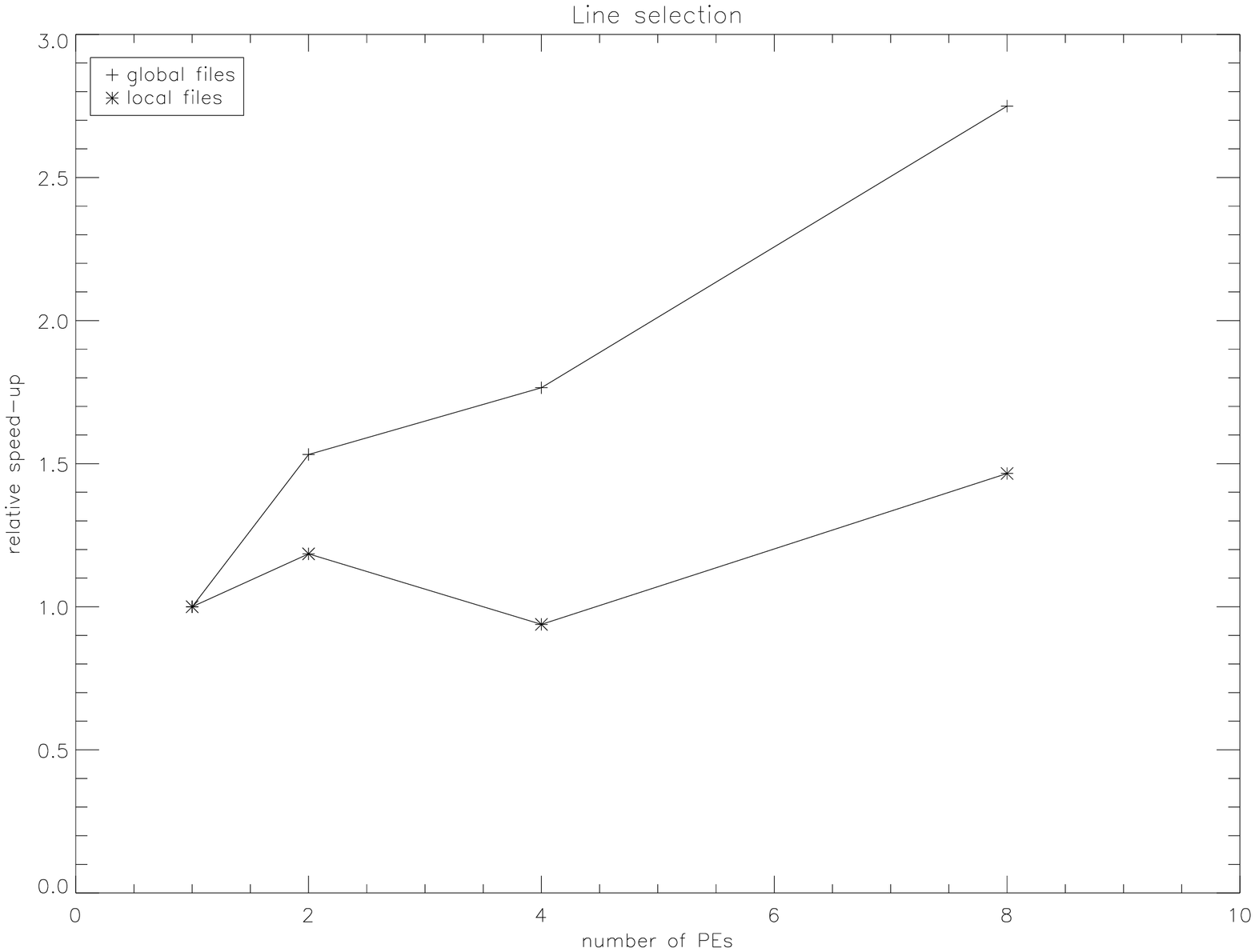,width=0.75\hsize,angle=90}
\caption[]{\label{ppro_line_sel}Results for the line selection 
process 
on the PPro/Solaris system. Times are given in wallclock seconds 
in the upper panel and as relative speedups for the lower panel.}
\end{figure}

The results for the line selection procedure on the 
PPro/Solaris system are shown in Fig.~\ref{ppro_line_sel}.
It is apparent from the figure that the GTF approach 
delivers higher relative speedups that translate into
smaller execution times for more than 2 PEs. For serial
(1PE) and 2PE parallel runs the LTF line selection 
is substantially faster than the GTF algorithm. The reason
for this behavior can be explained by noting that the 
access of the global files is done through NFS mounts
that use the same network as the MPI messages. Therefore,
$n-1$ PEs request different data blocks from the NFS
server (no process was run on the NFS server itself) 
and send their results to the I/O PE, which 
writes it out to the NFS server. In the LTF algorithm,
each PE reads a different input block from the NFS
server and then sends its results (around the ring) to all other PEs. 
Upon receiving data from its left neighbor, a PE
writes it to local disk. This means that the amount
of data streaming over the network can be as much
as twice as high for the LTF compared to the 
GTF algorithm. This increases the execution time
for the LTF approach if the network utilization is close 
to the maximum bandwidth. In this argument we have ignored 
the time required to write the data to local disks, which 
would make the situation worse for the LTF approach.

\begin{figure}[b]
\psfig{file=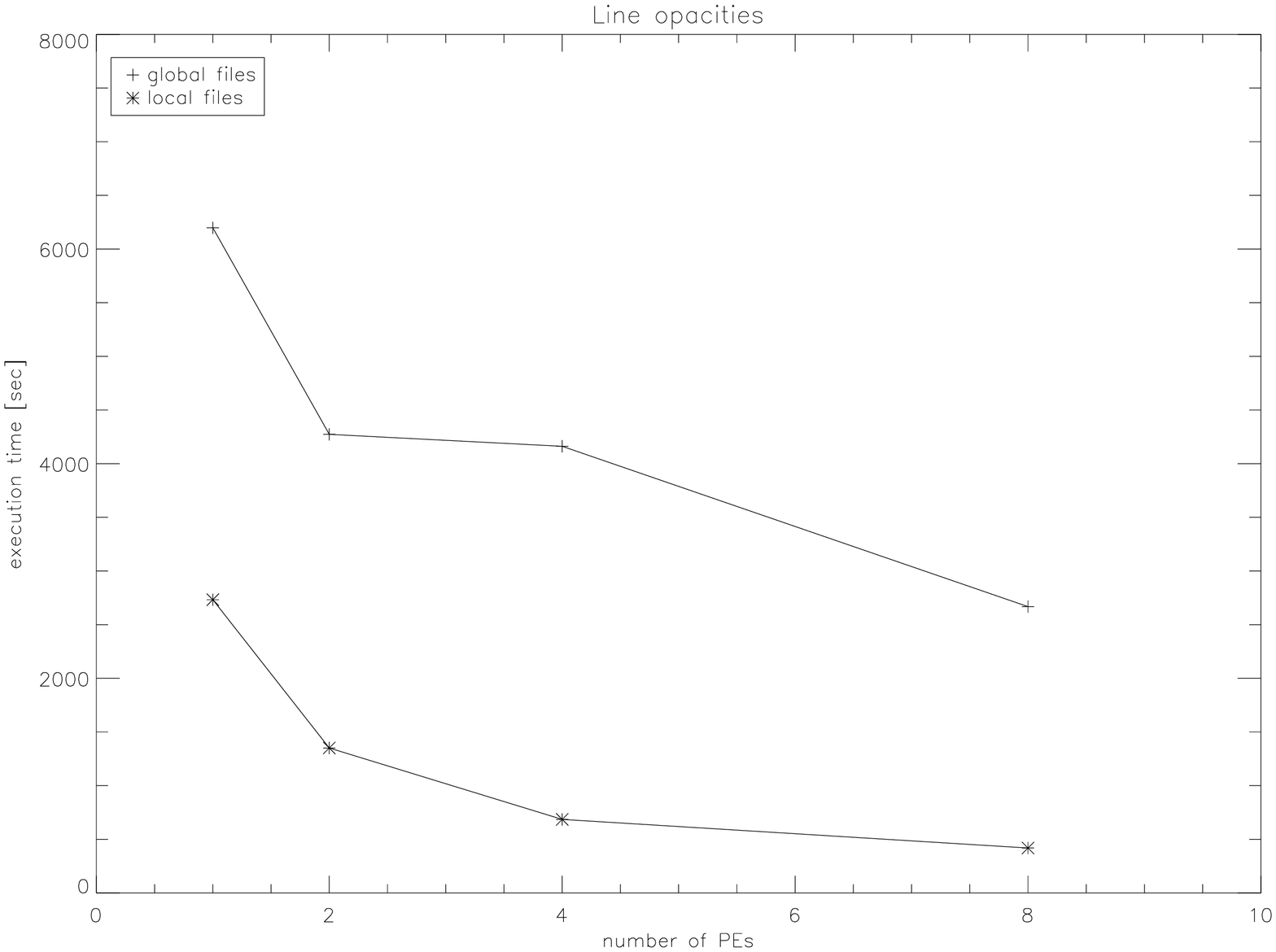,width=0.75\hsize,angle=90}
\psfig{file=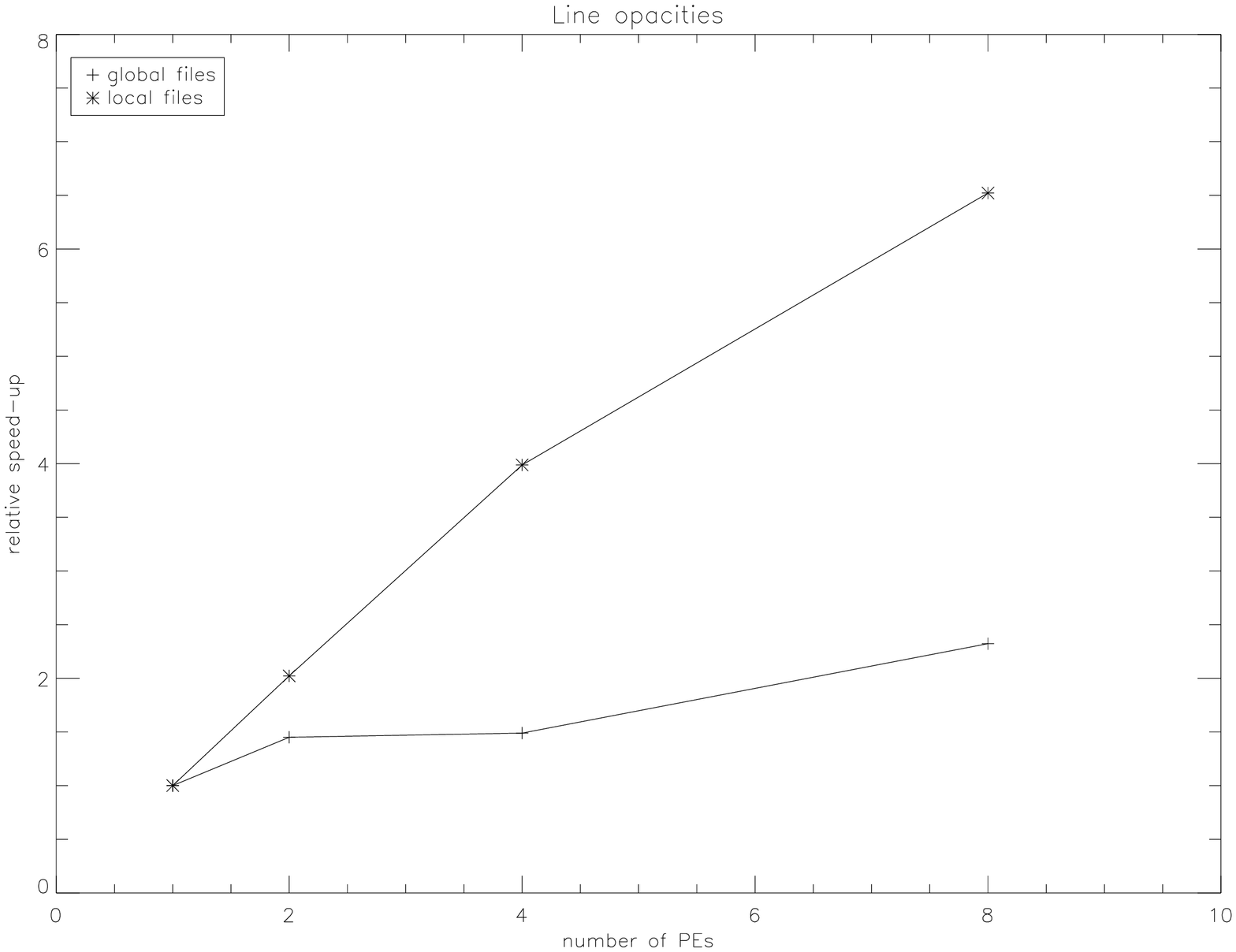,width=0.75\hsize,angle=90}
\caption[]{\label{ppro_line_op}Results for the line opacity 
calculations 
on the PPro/Solaris system. Times are given in wallclock seconds 
in the upper panel and as relative speedups for the lower panel.}
\end{figure}

The situation is very different for the calculation of the line opacities,
c.f.\ Fig.~\ref{ppro_line_op}. Now the LTF approach scales well (up to the
maximum of 8 available machines) whereas the GTF algorithm hardly scales to
more than 2 PEs. The absolute execution times for the LTF approach are up to a
factor of 4 smaller (more typical are factors around 2) than the corresponding
times for the GTF algorithm (the GTF run with 8PEs required roughly as much
execution time as the LTF run with 1PE!). The reason for this is clearly the
speed advantage of the local disk I/O compared to the NFS based I/O in the GTF
code.  If more PEs are used in the GTF line opacity approach, the network
becomes saturated quickly and the PEs have to wait for their data (the NFS
server itself was not the bottleneck).  The LTF approach will be limited by the
fact that as the number of PEs get larger, the efficiency of disk caching is
reduced and more physical I/O operations are required. Eventually this will
limit the scaling as the execution time is limited by physical I/O to local
disks.

\section{IBM SP system}

\begin{figure}[b]
\psfig{file=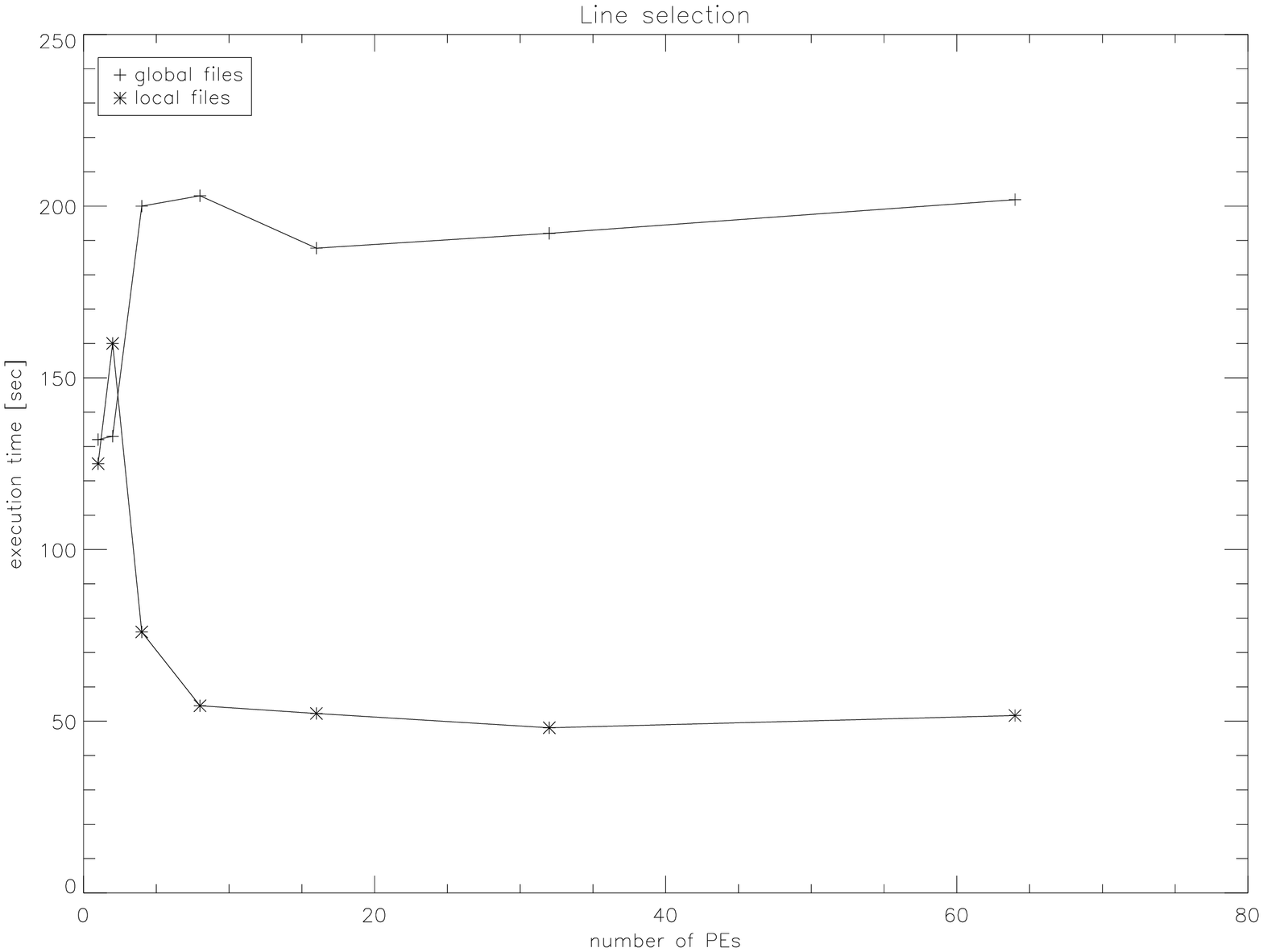,width=0.75\hsize,angle=90}
\psfig{file=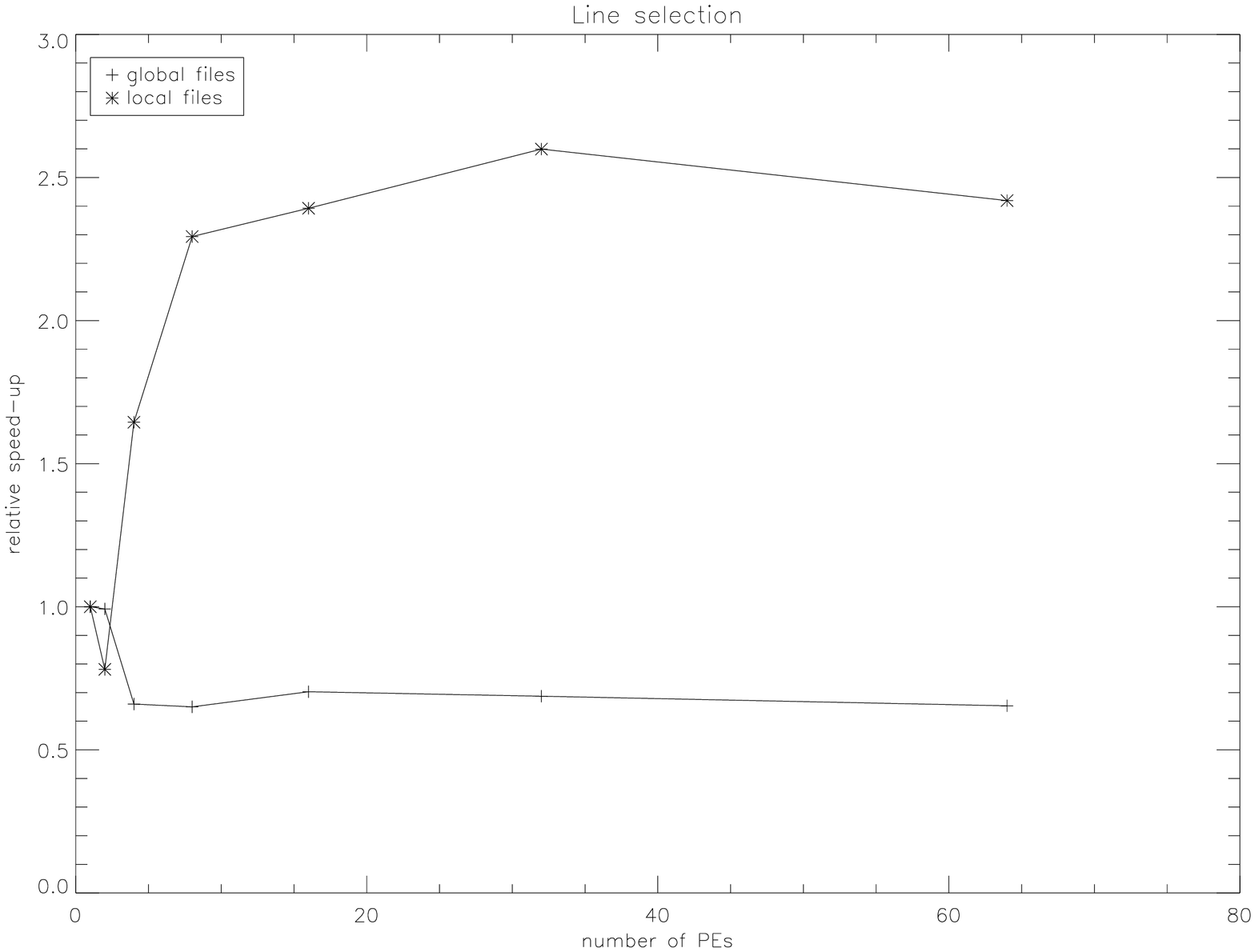,width=0.75\hsize,angle=90}
\caption[]{\label{ibm_line_sel}Results for the line selection 
process 
on the IBM SP system. Times are given in wallclock seconds 
in the upper panel and as relative speedups for the lower panel.}
\end{figure}

We ran the same (small) test on an IBM SP for comparison.  The tests were run
on a non-dedicated production system and thus timings are representative of
standard operation conditions and not optimum values.  The global files were
stored on IBM's General Purpose File System (GPFS), which is installed on a
number of system-dedicated I/O nodes replacing the NFS fileserver used on the
PPro/Solaris system. GPFS access is facilitated through the same ``switch''
architecture that also carries MPI messages on the IBM SP.  The results for the
line selection code are shown in Fig.~\ref{ibm_line_sel}. For the small test
the results are markedly different from the results for the PPro/Solaris
system. The LTF algorithm performs significantly better for all tested
configurations, however, scaling is very limited. The GTF code does not scale
well at all for this small test on the IBM SP.  This is due to the small size,
so that the processing is so fast (nearly 100 times faster than on the
PPro/Solaris system) so that, e.g., latencies and actual line selection calculations
overwhelm the timing. The IBM SP has a very fast switched communications
network that can easily handle the higher message traffic created by the LTF
code. This  
explains why the LTF line selection executes faster and scales better for
this small test on the IBM compared to the PPro/Solaris system.

\begin{figure}[b]
\psfig{file=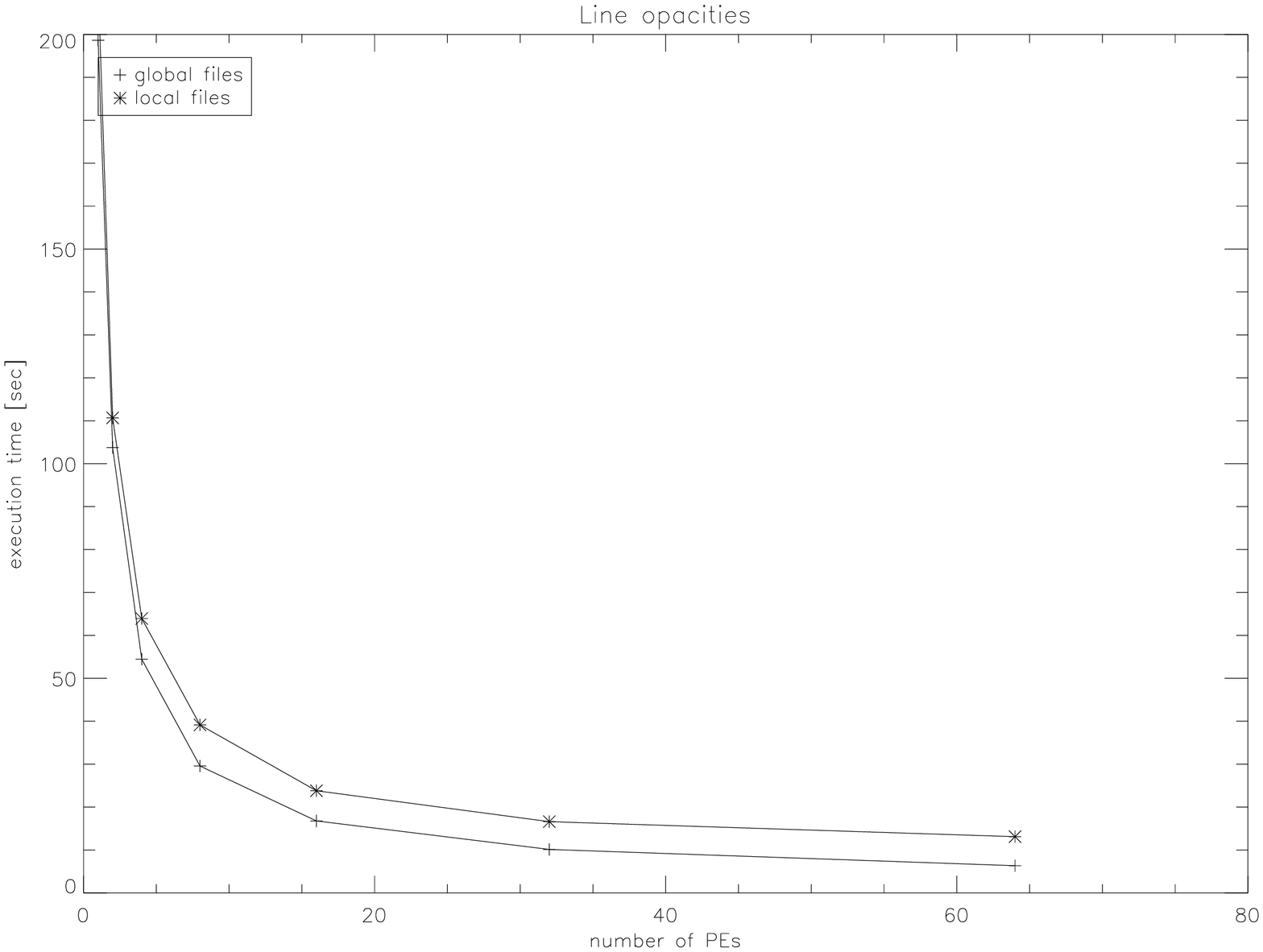,width=0.75\hsize,angle=90}
\psfig{file=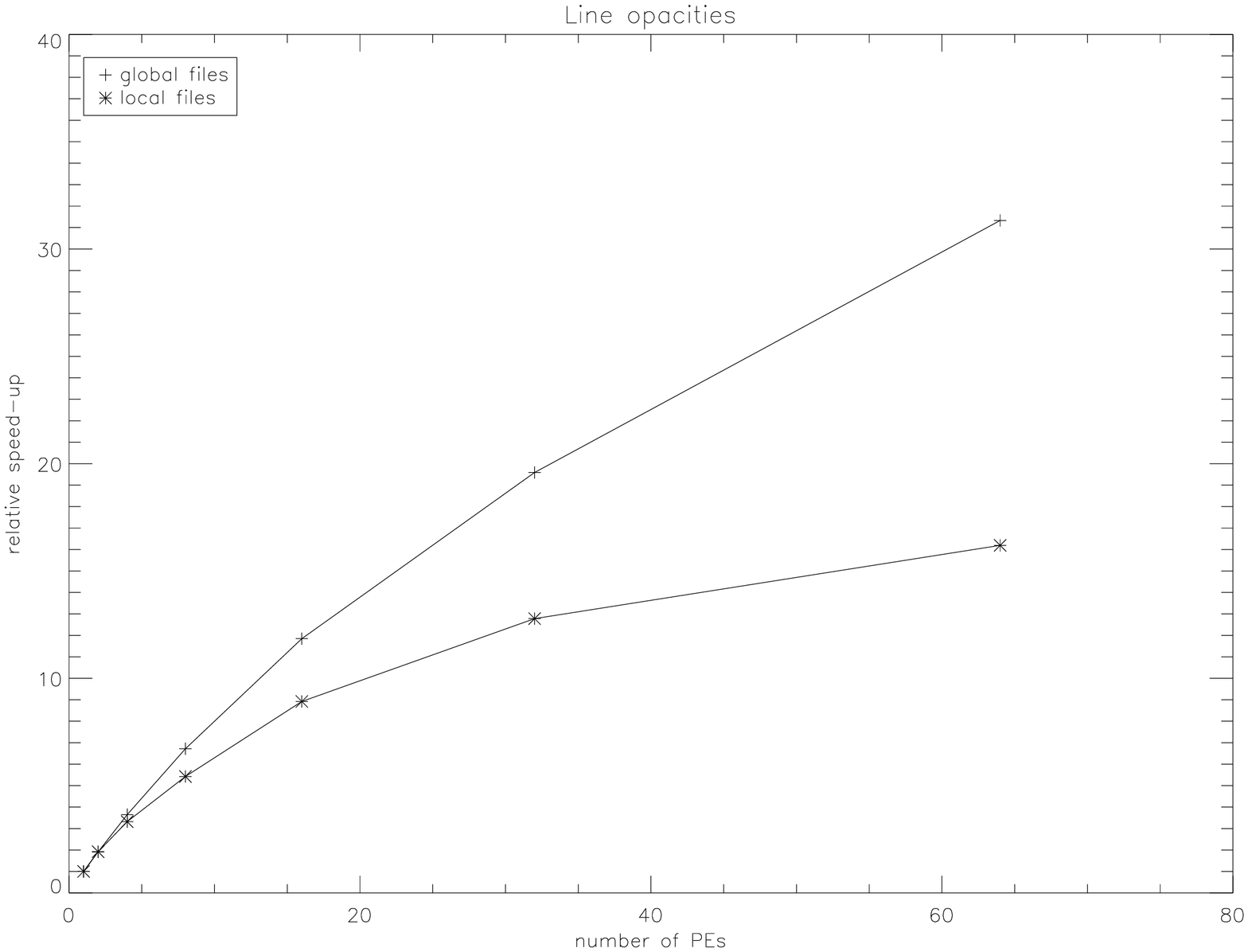,width=0.75\hsize,angle=90}
\caption[]{\label{ibm_line_op}Results for the line opacity 
calculations 
on the IBM SP system. Times are given in wallclock seconds 
in the upper panel and as relative speedups for the lower panel.}
\end{figure}

 The line opacity part of the test shown in Fig.~\ref{ibm_line_op}
performs distinctively different in the IBM SP compared to the 
PPro/Solaris system (cf.\ Fig.~\ref{ppro_line_op}). In contrast to 
the latter, the IBM SP delivers better performance for the GTF 
algorithm compared to the LTF code. The scaling of the GTF code is 
also significantly better than that of the LTF approach. This 
surprising result is a consequence of the high I/O bandwidth of the
GPFS running on many I/O nodes, the I/O bandwidth available
to GPFS is significantly higher than the bandwidth of
the local disks (including all filesystem overheads etc). The I/O nodes
of the GPFS can also cache blocks in their memory which can eliminate
physical I/O to a disk and replace access by data exchange over the 
IBM ``switch''. Note that the test was designed and run with parameters
set to maximize actual I/O operations in order to explicitly test
this property of the algorithms.

\begin{figure}[b]
\psfig{file=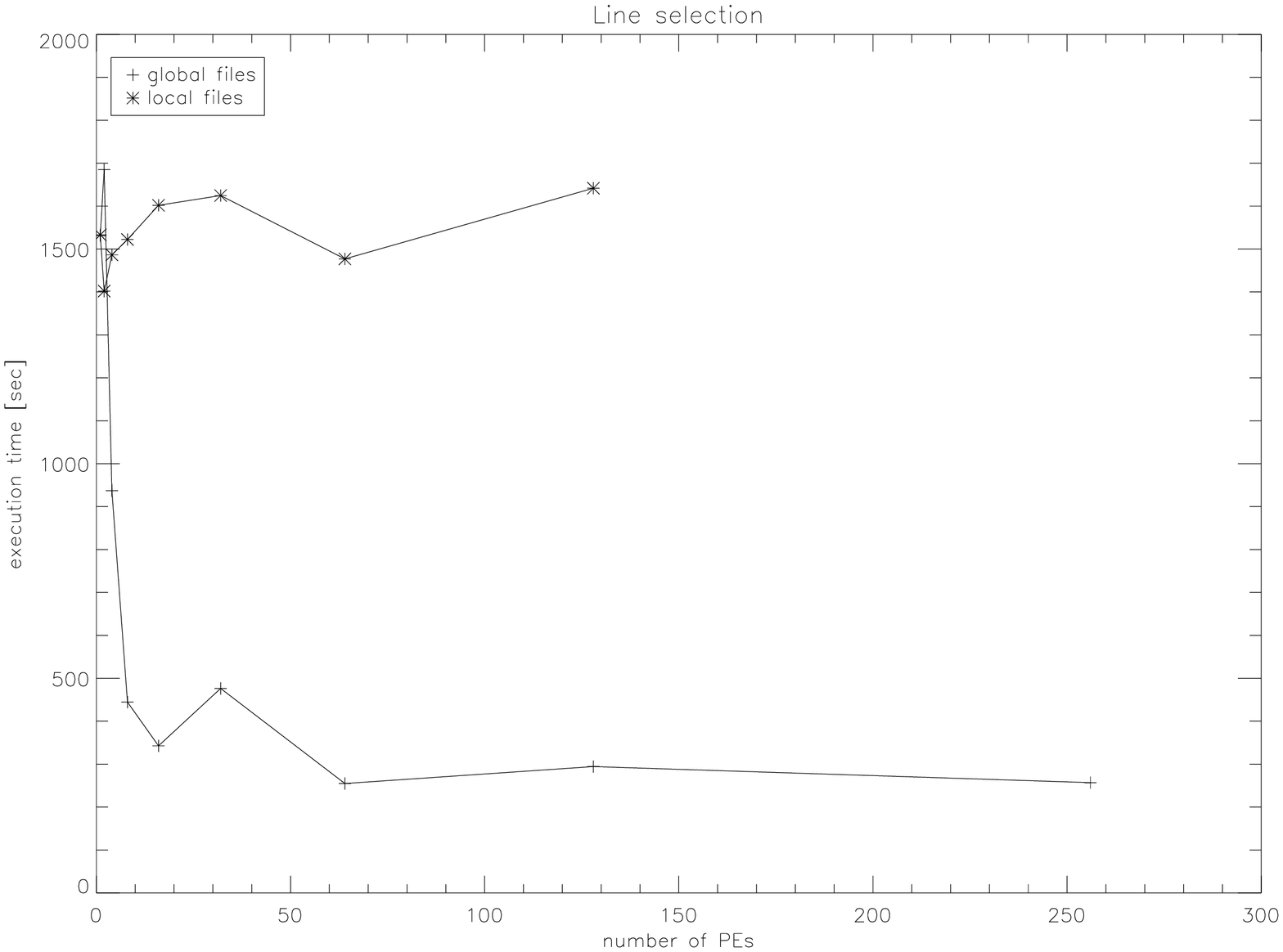,width=0.75\hsize,angle=90}
\psfig{file=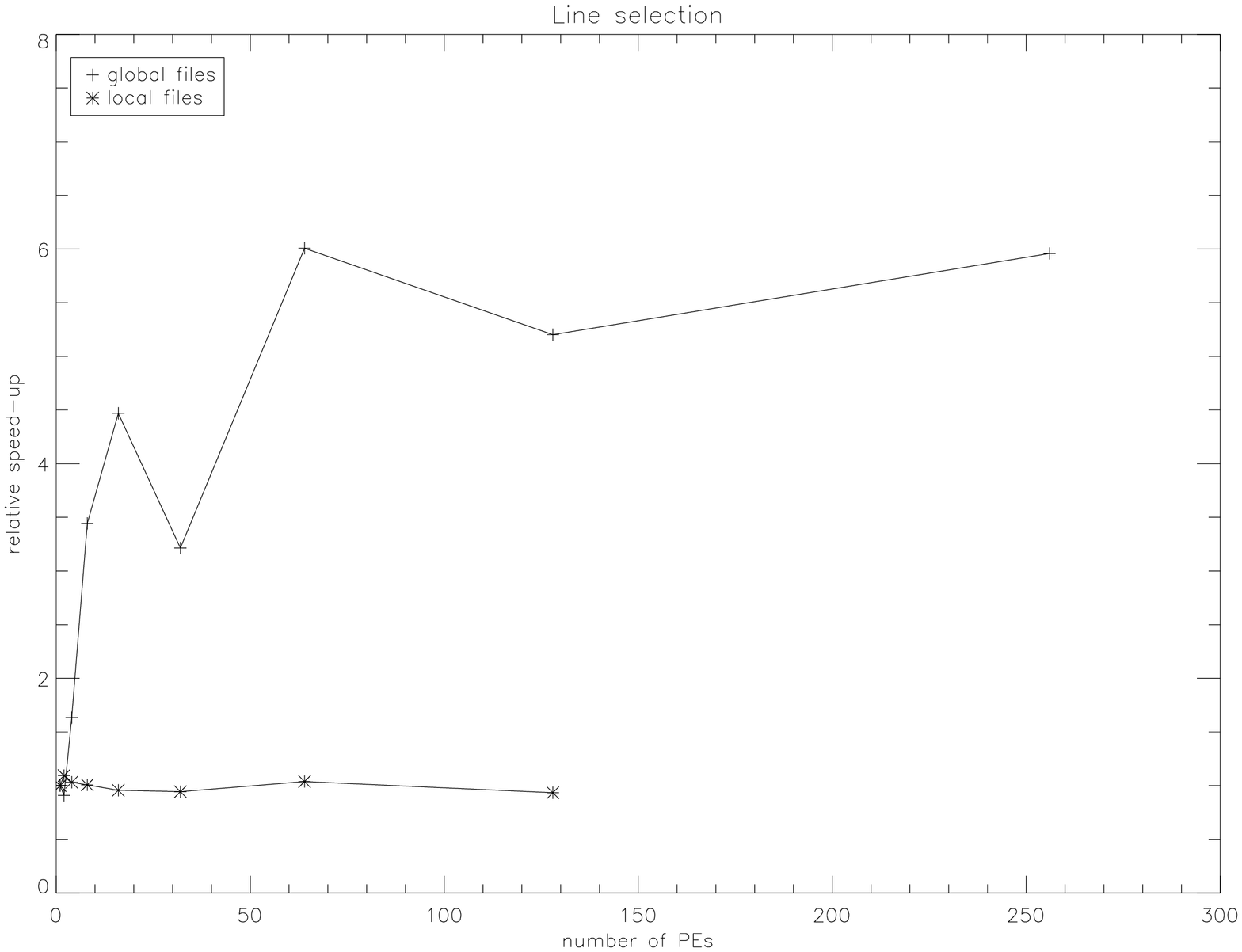,width=0.75\hsize,angle=90}
\caption[]{\label{ibm_large_line_sel}Results for the line selection 
process for the large test 
on the IBM SP system. Times are given in wallclock seconds 
in the upper panel and as relative speedups for the lower panel.}
\end{figure}

The results of the large test case, for which the input file
size is about 16 times bigger, are very different for the 
line selection, cf.\ Fig.~\ref{ibm_large_line_sel}. Now the 
GTF algorithm executes much faster (factor of 3) than the 
LTF code. This is probably caused by the larger I/O performance
of the GPFS that can easily deliver the data to all nodes and 
the smaller number or messages that need to be exchanged in the 
GTF algorithm. The drop of performance at 32PE's in the GTF line
selection run could have been caused by a temporary overload
of the I/O subsystem (these tests were run on a non-dedicated 
machine). 
In contrast to the previous tests, 
the LTF approach does not scale in this case.
This is likely caused
by the large number of relatively small messages that are 
exchanged by the PEs (the line list master database is the same 
as for the GTF approach, so it is read through the GPFS as in
the GTF case). This could be improved, e.g., by choosing larger
block sizes for data sent via MPI messages, however, this 
will have the drawback of more memory usage and larger messages
are more likely to block than small messages that can be stored
within the communication hardware (or driver) itself. 

\begin{figure}[b]
\psfig{file=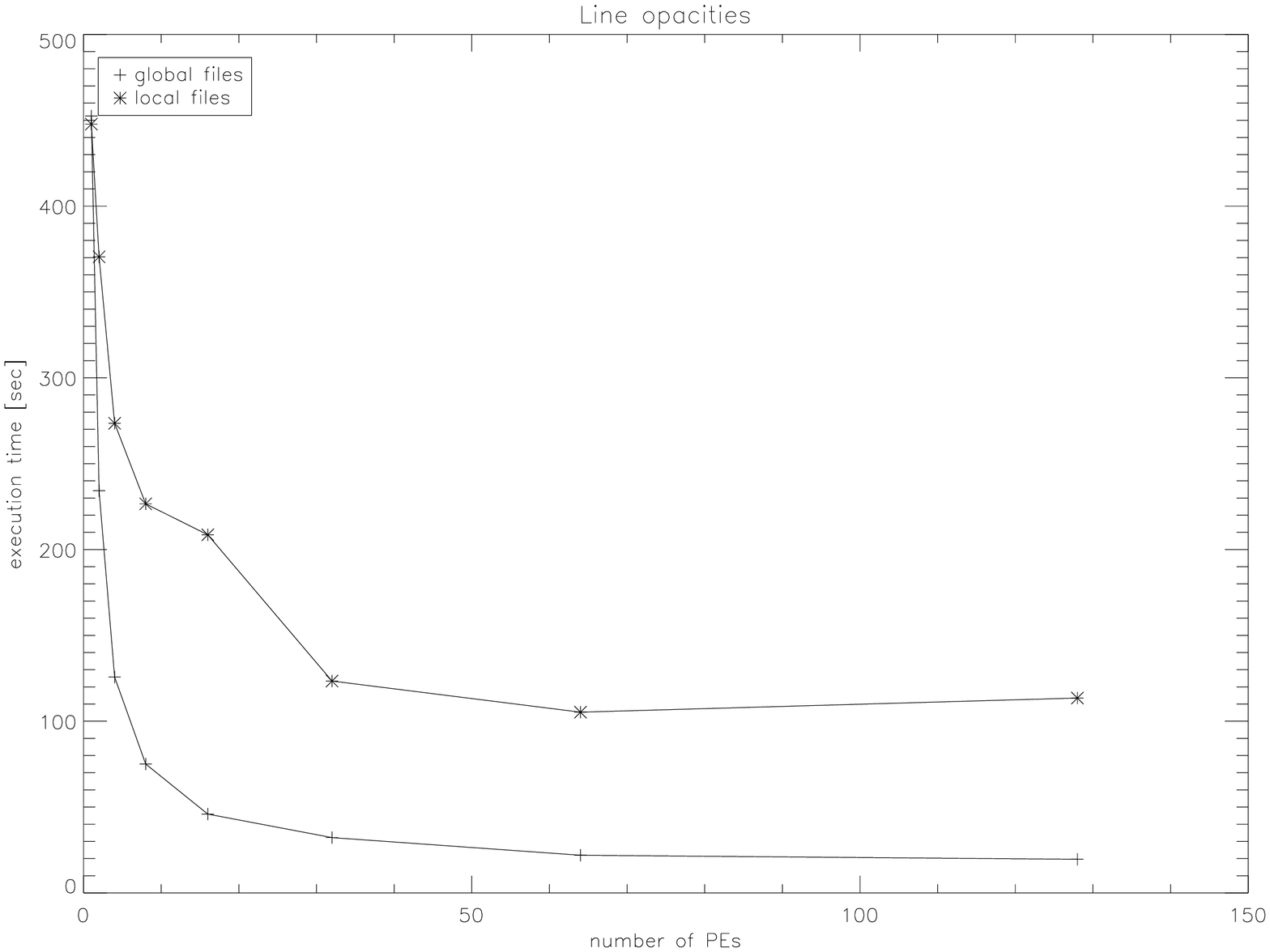,width=0.75\hsize,angle=90}
\psfig{file=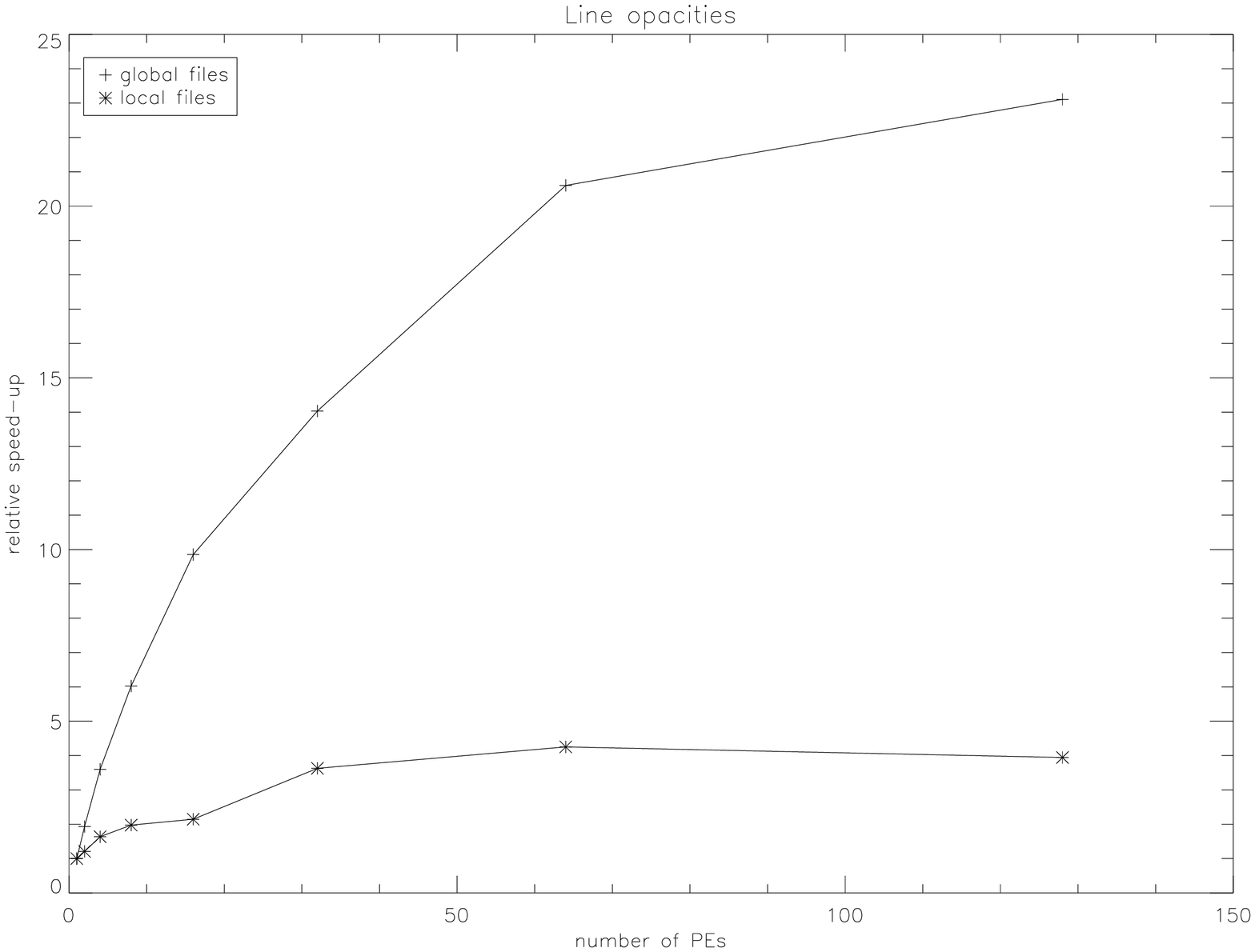,width=0.75\hsize,angle=90}
\caption[]{\label{ibm_large_line_op}Results for the line opacity 
calculations for the large test
on the IBM SP system. Times are given in wallclock seconds 
in the upper panel and as relative speedups for the lower panel.}
\end{figure}

The situation for the line opacities is shown in Fig.~\ref{ibm_large_line_op}.
The scaling is somewhat worse due to increased physical I/O (the temporary
files are about 16 times larger as in the small test case).  This is
more problematic for the LTF approach which scales very poorly for larger
numbers of PEs as the maximum local I/O bandwidth is reached far earlier than
for the GTF approach. This is rather surprising as conventional
wisdom would favor local disk I/O over global filesystem I/O on parallel 
machines. Although this is certainly true for farms of workstations or 
PCs, this is evidently not true on high-performance parallel computers 
with parallel I/O subsystems. 
 
\section{Summary and Conclusions}

In this paper we have discussed two algorithms for 
parallel spectral line selection and opacity calculations 
useful for direct opacity sampling models of stellar atmospheres.
The GTF algorithm uses global temporary files to store the
list of selected lines, whereas the LTF algorithm stores 
the scratch files on local disks. These two methods show
very different performance characteristics on different
parallel systems. On a PC cluster, each processing element
(PE) being a PC with its own local disk space and operating 
system networked with standard IP based Ethernet, the LTF algorithm
has disadvantages in the line selection procedure for
larger numbers of PEs ($\ge 4$), probably
due to the higher demand put on the communication between nodes
using the MPI library, but is slightly faster than the GTF approach
if the number of PEs is smaller.
However the LTF code produces far faster and 
better scaling line opacity calculations, which will be the 
more costly part of a typical atmosphere model run (the line selection
is usually required only once at the beginning of a model 
calculation). 

On the common IBM SP parallel supercomputers the situation changes
significantly. On this machine, the small test runs so fast that
the timing is dominated by side effects. In the  large test case, the
GTF line selection performs and scales far better than the LTF code.
This surprising result is caused by the presence of a parallel 
filesystem (GPFS) on the IBM that dramatically improves performance
of global I/O compared to local disk I/O. The GPFS also boosts the
performance for the GTF code for the (in practical applications more
important) line opacity calculations, for both the small and the large
test cases.

Variations of the algorithms can be constructed, e.g., it is possible to store
the master line database on each PE individually and thus totally remove global
I/O to a single master line list (this requires enough local disk space to
store both the master and temporary databases). Other improvements are
possible, e.g., optimization of the I/O blocksize for each type of machine.
However, these optimizations are system dependent (and also depend on the load
of the machine in general) and thus are not discussed here.

The algorithms and the results show that parallel computing 
can lead to dramatic speed improvements in stellar atmosphere
calculations but also that different algorithms are required 
for different types and capabilities of parallel machines.
The speed improvements can then be used to develop physically
more complex and detailed models (e.g., including massive 
NLTE calculations with line blanketing, models for M dwarfs 
with possibly billions of molecular spectra lines or detailed 
models for stellar winds and for hot stars  with radiative
levitation modeling including a large number of elements and 
ionization stages). This approach sends us one step further to
a better physical understanding of stars and their spectra.

Both algorithms we have described here are applicable to a wide range of
problems that have data requirements that are larger than the available memory
and thus need to perform out-of-core calculations. They can be trivially
adapted to any case in which a large amount of shared data has to be utilized
by a number of processors simultaneously and where it is not easy to use, e.g.,
a domain-decomposition approach to allow each processor to use only a distinct,
smaller subset of the data.  If the exchange of data can be arranged in a
ring-like topology and the communication network of the parallel computer used
is fast, then the LTF algorithm should be efficient, however, if the machine
has a fast parallel filesystem, then the GTF approach is both simpler to
implement and more efficient.

\acknowledgments 

This work was supported in part by NSF grant AST-9720704, NASA ATP grant NAG
5-8425 and LTSA grant NAG 5-3619, as well as NASA/JPL grant 961582 to the
University of Georgia and in part by NSF grants AST-97314508, by NASA grant
NAG5-3505 and an IBM SUR grant to the University of Oklahoma.  This work was
supported in part by the P\^ole Scientifique de Mod\'elisation Num\'erique at
ENS-Lyon.  Some of the calculations presented in this paper were performed on
the IBM SP2 of the UGA UCNS, on the IBM SP ``Blue Horizon'' of the San Diego
Supercomputer Center (SDSC), with support from the National Science Foundation,
and on the IBM SP of the NERSC with support from the DoE.  We thank all these
institutions for a generous allocation of computer time.

\clearpage

\bibliography{yeti,opacity,mdwarf,radtran,general,opacity-fa,mdwarf-fa}

\clearpage
\section{Figures}

\end{document}